\documentclass[conference]{IEEEtran}
\IEEEoverridecommandlockouts
\usepackage{cite}
\usepackage{amsmath,amssymb,amsfonts}
\usepackage{algorithmic}
\usepackage{graphicx}
\usepackage{textcomp}

\usepackage[table,xcdraw]{xcolor}
\usepackage{multirow}
\usepackage{url}
\usepackage{booktabs}
\usepackage{comment}
\usepackage{algorithm}
\def\BibTeX{{\rm B\kern-.05em{\sc i\kern-.025em b}\kern-.08em
    T\kern-.1667em\lower.7ex\hbox{E}\kern-.125emX}}
\begin{document}
\title{CEFW: A Comprehensive Evaluation Framework for Watermark in Large Language Models}

\makeatletter
\newcommand{\linebreakand}{%
  \end{@IEEEauthorhalign}
  \hfill\mbox{}\par
  \mbox{}\hfill\begin{@IEEEauthorhalign}
}
\makeatother

\author{\IEEEauthorblockN{Shuhao Zhang$^{1}$\thanks{\(^\dag\) Corresponding Author. }, Bo Cheng$^{1\dag}$, Jiale Han$^2$, Yuli Chen$^1$, Zhixuan Wu$^1$, Changbao Li$^3$, Pingli Gu$^3$}
\IEEEauthorblockA{
\textit{$^1$State Key Laboratory of Networking and Switching Technology, Beijing University of Posts and Telecommunications} \\
\textit{$^2$Hong Kong University of Science and Technology} \\
\textit{$^3$BigData R$\&$D Center, North China Institute of Computing Technology} \\
\{2020111429,chengbo,chenyuli,wzxmogu\}@bupt.edu.cn, jialehan@ust.hk, 
\{lichangbao\_1, gpl98\}@163.com}
}
\maketitle

\begin{abstract}
Text watermarking provides an effective solution for identifying synthetic text generated by large language models. However, existing techniques often focus on satisfying specific criteria while ignoring other key aspects, lacking a unified evaluation. To fill this gap, we propose the Comprehensive Evaluation Framework for Watermark (CEFW), a unified framework that comprehensively evaluates watermarking methods across five key dimensions: ease of detection, fidelity of text quality, minimal embedding cost, robustness to adversarial attacks, and imperceptibility to prevent imitation or forgery. By assessing watermarks according to all these key criteria, CEFW offers a thorough evaluation of their practicality and effectiveness. Moreover, we introduce a simple and effective watermarking method called Balanced Watermark (BW), which guarantees robustness and imperceptibility through balancing the way watermark information is added. Extensive experiments show that BW outperforms existing methods in overall performance across all evaluation dimensions. We release our code to the community for future research\footnote{\url{https://github.com/DrankXs/BalancedWatermark}}.
\end{abstract}


\section{Introduction}
\label{sec:intro}

\begin{figure}[t]
  \centering
  \includegraphics[width=0.95\columnwidth]{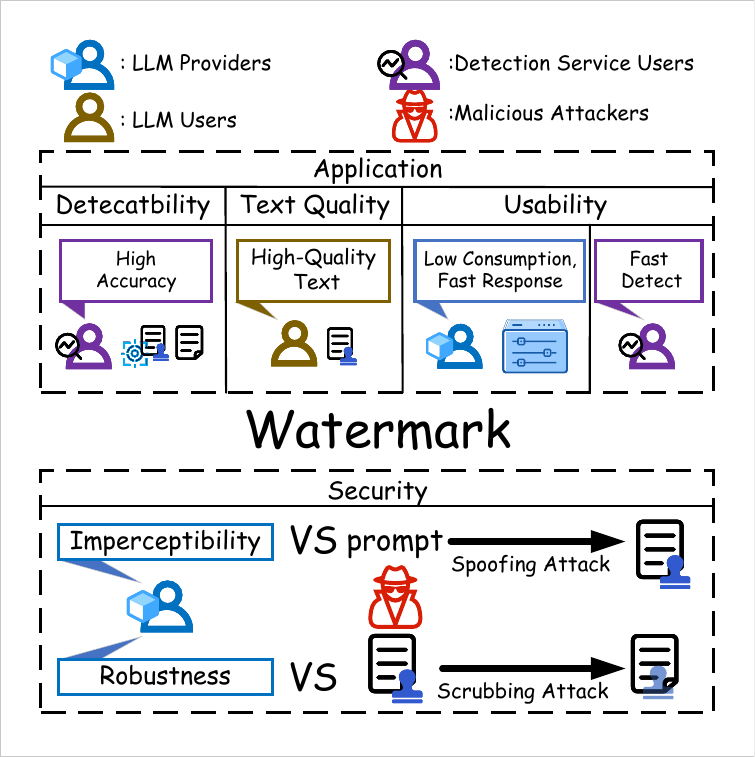}
  \vspace{-4mm}
  \caption{Watermarks have three distinct audiences: LLM providers, LLM users, and detection service users. They necessitate watermarks to have detectability, text quality, and usability in terms of application. Concurrently, LLM providers must ensure the imperceptibility and robustness of watermarks in terms of security to counteract scrubbing attacks and spoofing attacks by malicious attackers. }
  \label{fig:fiveCharacteristics}
\end{figure}
With the rapid development of large language models (LLMs) \cite{LMDGPT4, Modell3llama3modelcard}, text generated by LLMs increasingly resembles human-generated text and gradually fills all parts of our lives. 
This trend presents several potential threats, including hallucinations \cite{LMGalkaissi2023artificial}, misinformation generation \cite{misinformation2:conf/www/ZhangSDL24}, and malicious use \cite{LMDGPT4, museeditorials2023tools}. 
Therefore, detecting text generated by LLMs has become an emerging and critical issue.

Digital watermark \cite{WMNaturlWatermarking01,KGW} is a promising approach for detecting LLM-generated text. 
The approach embeds watermark information into the text and determines whether the text is generated by the LLM by detecting the watermark information. 
As shown in Fig. \ref{fig:fiveCharacteristics}, an effective watermark should meet both applicability and security requirements, demonstrating the following five characteristics:
(1) \textbf{Detectability}: The ability to accurately distinguish between watermarked and non-watermarked text; 
(2) \textbf{Text Quality}: The quality of watermarked texts should not significantly be compromised; 
(3) \textbf{Usability}: The time and resource consumption incurred by the addition and detection of a watermark should be acceptable; 
(4) \textbf{Robustness}: The watermark should remain detectable even when the watermarked text is subjected to scrubbing attacks; 
(5) \textbf{Imperceptibility}: The watermark should not be cracked by spoofing attacks, ensuring that attackers cannot generate legitimate watermark text.

Previous watermark studies \cite{KGW,WMSW24,WMUNIW23,WMInvisiblewm23,WMsemstamp23,WMAdaptiveWatermark,WMEWD24} focus only on part of the necessary watermark characteristics without considering the whole picture for their watermarks. 
KGW \cite{KGW} emphasizes detectability and robustness, but neglects usability and imperceptibility when evaluation. 
Unigram Watermark (UNIW) \cite{WMUNIW23} aims to enhance robustness, however, the imperceptibility neglected by UNIW is an obviously shortcoming. 
Although these watermarks are proven to be effective for the characteristics they focus on, the lack of evaluation of the other characteristics makes them unreliable, thereby hindering their practicality. 
The ``barrel effect'' illustrates that the capacity of a barrel is determined by its shortest plank, highlighting the idea that the weakest component often limits the overall performance or potential of a system. 
Therefore, a uniform and complete watermark evaluation framework is highly important.

In this paper, we propose the Comprehensive Evaluation Framework for Watermark in LLMs (CEFW) to obtain a comprehensive evaluation for the watermark. 
CEFW selects watermark metrics for each characteristic and calculates them to obtain the corresponding score. 
By weighting five characteristic scores, CEFW offers a comprehensive score for the evaluated watermark. 
We also notice that the way for watermark information adding is over-dynamic in KGW, and over-static in UNIW. 
Therefore, we propose Balanced Watermark (BW), it is designed to balance the way watermark information is added. 
BW integrates the advantages of both dynamic and static aspects to obtain a higher comprehensive score. 
Our experiment proves that BW not only has the best comprehensive performance, but also has no obvious shortcomings. 

Our main contributions are as follows: 
\begin{itemize} 
  \item We propose the Comprehensive Evaluation Framework (CEFW) for Watermark in LLM from five key dimensions, which first standardizes the watermark evaluation process and provides a comprehensive score for watermark. 
  \item We introduce the Balanced Watermark (BW), which balances the way of adding watermark information to avoid being over-dynamic or over-static, offering a well-rounded and practical solution. 
  \item We carry out extensive experiments to obtain the comprehensive score of different watermarks using the CEFW framework, and experimental results demonstrate that BW achieves the highest comprehensive performance. 

\end{itemize}

\section{Related Work}

\subsection{LLM-Generated Text Watermark}
In order to distinguish between texts generated by models and those composed in natural language, 
some scholars try to find a more accurate detector \cite{DEGLTR19,DEDetectGPT23}, while others decided to tackle the problem at the source by adding watermarks to the LLM-generated text.
Watermarks for LLM-generated text can be categorized into three types based on different embedding phrases:
A-prior watermarks modify parameters of LLM \cite{BDadi2018turning,BDDBLP:conf/acl/PengYWWZLJXSX23}; 
generating watermarks change the output probabilities of LLM \cite{KGW,WMEWD24,WMSW24,WMUNIW23,WMUnbiasedWatermark}; 
plain text watermarks edit existing texts \cite{WMLexical22,WMWordImportance}. 
Generating watermarks emerge as a focal point of current watermark research \cite{WMSIR:conf/iclr/LiuPHM024,WMAdaptiveWatermark,WMUnforgeable-DBLP:conf/iclr/LiuPH0WKY24}. 
It does not require fine-tuning LLM, and effectively preserves the original inference capabilities of the LLM.

\subsection{Watermark Evaluation}
Most researchers, when proposing a new watermark, only evaluate detectability, text quality, and robustness \cite{KGW,WMUNIW23,WMSW24}. 
The measurement of detectability typically employs binary classification metrics, while text quality is represented by perplexity and rouge score. 
For robustness, various researchers employ different scrubbing attacks to remove the watermark from the text and analyze changes in detectability to assess robustness.
The most effective approach is DIPPER \cite{dipperLFQA:conf/nips/KrishnaSKWI23}. 
Liu et al.\cite{WMSIR:conf/iclr/LiuPHM024} take into account the usability and imperceptibility required by the watermark. 
They evaluate usability by comparing changes in generation speed, but overlook the additional memory consumption caused by the watermark. 
Regarding imperceptibility (referred to as security robustness), Liu et al.\cite{WMSIR:conf/iclr/LiuPHM024} apply the spoofing attack proposed by \cite{AT-spoofingDBLP:journals/corr/abs-2303-11156} and test the success rate of the attack to evaluate imperceptibility.

\section{Comprehensive Evaluation Framework for Watermark}
In this section, we first analyze why the five characteristics are necessary for watermark evaluation.
Subsequently, we present the overall process of the Comprehensive Evaluation Framework for Watermark in LLMs (CEFW).

\subsection{Role Analysis}
The five characteristics are necessary for watermark evaluation due to the requirements of actual LLM service. 

Actual LLM service with the watermark primarily involves four roles: LLM provider, LLM user, detection service user, and malicious attacker. 

At the application level, LLM providers should consider the impact of watermarks on memory consumption and response speed during deployment, which refers to \textbf{usability}. 
When using LLM, LLM users expect more accurate responses, signifying higher \textbf{text quality}. 
Detection service users desire more accurate and swift detection of watermarked text, necessitating better \textbf{detectability} and \textbf{usability} of the watermark. 

From a security perspective, malicious attackers attempt two types of attack method: scrubbing attack and spoofing attack.
Scrubbing attacks modify the watermarked text with the aim of removing the watermark while preserving the meaning of the text generated by LLM. 
LLM providers require watermarks with higher \textbf{robustness} to defend against scrubbing attacks.
Spoofing attacks involve unauthorized cracking of the watermark and simulating watermarked text, which can undermine the credibility of target watermark. 
To resist spoofing attacks, watermarks are required to demonstrate greater \textbf{imperceptibility}. 
\label{role analyse}

\subsection{Overall Process}

We first define a few symbols: 
$Y$ is the text generated by LLM without watermark; 
$\hat{Y}$ is the watermarked text; 
$X$ is the prompt for generating $Y$ and $\hat{Y}$; 
$A(\cdot)$ is the text generated by the attack algorithm; 
$M(\cdot)$ is the metric function to calculate the metric value; 
$S$ is the score obtained after normalization. 

\paragraph{Detectability}
We select Area Under the Receiver Operating Characteristic Curve (AUCROC) for detectability evaluation. 
AUCROC integrates the information of True Positive Rate (TPR) and False Positive Rate (FPR) to comprehensively evaluate the detectability of watermark.
The worst AUCROC value is 0.5, which indicates a random classification of texts, while the best is 1. 
CEFW normalizes it at the following: 
\begin{equation}
    \label{eq:aucroc}
    S_{D} = \frac{M_{ACUROC}(Y, \hat{Y})-0.5}{1-0.5}
\end{equation}
$S_D$ is the detectability score. 

\paragraph{Text Quality}
We utilize perplexity (PPL) as the metric for text quality, which is the most popular in watermark evaluation. 
The advantage of PPL is versatile. 
PPL does not lose its evaluation function as ROUGE does due to changes in the generation task. 
CEFW obtains text quality score $S_T$ by the following: 
\begin{equation}
    \label{eq:ppl}
    S_{T} = \frac{M_{PPL}(\hat{Y})-2 M_{PPL}(Y)}{M_{PPL}(Y)-2M_{PPL}(Y)}
\end{equation}
We believe that evaluating the impact of watermark on text quality depends on the degree of degradation it brings to LLM. 
Eq. \eqref{eq:ppl} represents the best watermark will not degrade the LLM, while the worst one degrades it by up to twofold, indicating twice $M_{PPL}(Y)$. 

\paragraph{Usability}
Usability refers to the impact of the watermark on the normal use of LLM services and the efficiency of watermark detection.
In terms of time, we evaluate the Generate Time and Detect Time. 
As for memory aspect, we evaluate the Memory Cost after loading LLM and watermark. 
Like PPL, Memory Cost and Generate Time need to compare the change of metrics before and after adding the watermark to evaluate the impact of the watermark, they can utilize the same normalization method as Eq. \eqref{eq:ppl}. 
Scores $S_{MC}$ and $S_{GT}$ are obtained by normalizing the metric values of Memory Cost and Generate Time, respectively.
CEFW regards Detect Time as the time to detect a piece of text and sets the upper bound to 0 second and the lower bound to 1 second. 
We define the Detect Time score $S_{DT}$ as: 
\begin{equation}
    \label{eq:detectTime}
    S_{DT} = \frac{M_{DetectTime}(\hat{Y}) - 1}{0 - 1}
\end{equation} 
We average $S_{MC}$, $S_{GT}$ and $S_{DT}$ to obtain the usability score $S_U$.

\paragraph{Robustness}
Robustness is the ability of a watermark to remain detectable following a scrubbing attack. 
When the watermark has the best robustness, it will maintain the original detectability after a scrubbing attack. 
We compare the detectability metric value of watermarked text before and after scrubbing attack. 
CEFW selects AUCROC as the detectability metric and DIPPER as the scrubbing attack method. 
DIPPER is a language model to avoid watermark detection by paraphrasing text. 
The final robustness score $S_R$ is formalized as: 
\begin{equation}
    \label{eq:dipper}
    S_{R} = \frac{M_{AUCROC}(Y, A_{DIPPER}(\hat{Y})) - 0.5}{M_{AUCROC}(Y, \hat{Y}) - 0.5}
\end{equation}

\paragraph{Imperceptibility}
Imperceptibility is evaluated by detecting the existence of a watermark in the text generated by spoofing attacks. 
When spoofing attacks are most effective, they generate watermarked text that can be perfectly detected. 
We utilize AUCROC as the detectability metric and STEAL as the spoofing attack. 
STEAL score $S_{STEAL}$ is defined as: 
\begin{equation}
    \label{eq:steal}
    S_{STEAL} = \frac{M_{AUCROC}(Y, A_{STEAL}(X)) - 1}{0.5 - 1}
\end{equation}

STEAL statistically analyzes the n-gram frequency differences between non-watermarked text and watermarked texts to crack watermark. 
To achieve a more effective attack, CEFW performs four STEAL attacks by varying the value of n from 1 to 4 in the n-gram frequency analysis, as the watermark is unknown to the attackers. 
After implementing the spoofing attack procedure, CEFW can obtain four STEAL scores. 
To obtain more realistic imperceptibility scores through four STEAL scores, we conceive two scenarios for spoofing attack: 
1) Non-Authorization (N.A.): Malicious attacker lacks authorization to access the watermark detection service provided by LLM providers. 
2) Authorization (A.): Malicious attacker has unrestricted access to detect watermarked texts. 

Under N.A., malicious attacker lacks the capability to discern which spoofing attack is the most successful, thus can only make a random selection.
Under A., malicious attacker can easily select the most effective type of spoofing text generation models.

CEFW selects scenario A., where the imperceptibility score $S_I$ is determined as the minimum among the four $S_{STEAL}$.

\paragraph{Comprehensive} 
CEFW calculates a comprehensive score by weighting all five characteristic scores. 
The weights assigned are $\frac{1}{6}$ for detectability, $\frac{1}{6}$ for text quality, $\frac{1}{6}$ for usability, $\frac{1}{4}$ for robustness, $\frac{1}{4}$ for imperceptibility, 
Our weight assignment criterion is that applicability and security are equally important, and each characteristic under them has equal importance. 

\subsection{Flexible Design}

CEFW allows for the customization of metrics and their corresponding weights for each characteristic.
The available watermark metrics are detailed in the Appendix \ref{app:CEFW}.

\section{Balanced Watermark}
\label{sec:BW}
\subsection{Original Watermark}
Give a prompt $X=\{x_{1}, x_{2}, ..., x_{|X|}\}$, an LLM with parameters $\theta$ can generate a response $Y=\{y_{1}, y_{2}, ..., y_{|Y|}\}$. 
We can formula the probability distribution of the $i$-th token $y_i$ as:
\begin{equation}
    \label{eq:generate}
    P(y_i)=P_\theta(y_i|X, Y_{<i})
\end{equation}
To obtain $P(y_i)$, the LLM generates a logit value $l^{(i)}_k$ for each token $k$ in the vocabulary $\mathcal{V}$ at generation time. 

Both KGW and UNIW add watermark information by modifying $l^{(i)}_k$ at each generation step. 
They utilize corresponding \textbf{partition functions} to partition $\mathcal{V}$ into a green list $\mathcal{G}$ and a red list $\mathcal{R}$, and the logits of green tokens are increased by a positive constant $\delta$. 
The watermark probability $p^{(i)}_k$ can be formalized as follows:
\begin{equation}
    \label{eq:addlogitsbias} 
    p^{(i)}_{k}=
    \begin{cases}
        \frac{exp(l^{(i)}_k + \delta )}{\sum_{j\in \mathcal{R}} exp(l_j^{(i)} )  + \sum_{j \in \mathcal{G}} exp(l_j^{(i)} + \delta)}, &k \in \mathcal{G} \\
        \frac{exp(l^{(i)}_k)}{\sum_{j\in \mathcal{R}} exp(l_i^{(t)} ) + \sum_{j \in \mathcal{G}} exp(l_j^{(i)} + \delta)}, &k \in \mathcal{R}
    \end{cases}
\end{equation}

This results in an increased number of green tokens in watermarked text. 
We can calculate the number of green tokens in the given sentence, and utilize the z-statistic as the criterion to determine whether the watermark exists. 

\subsection{Partition Function}
\begin{figure}[t]
  \centering
  \includegraphics[width=0.95\columnwidth]{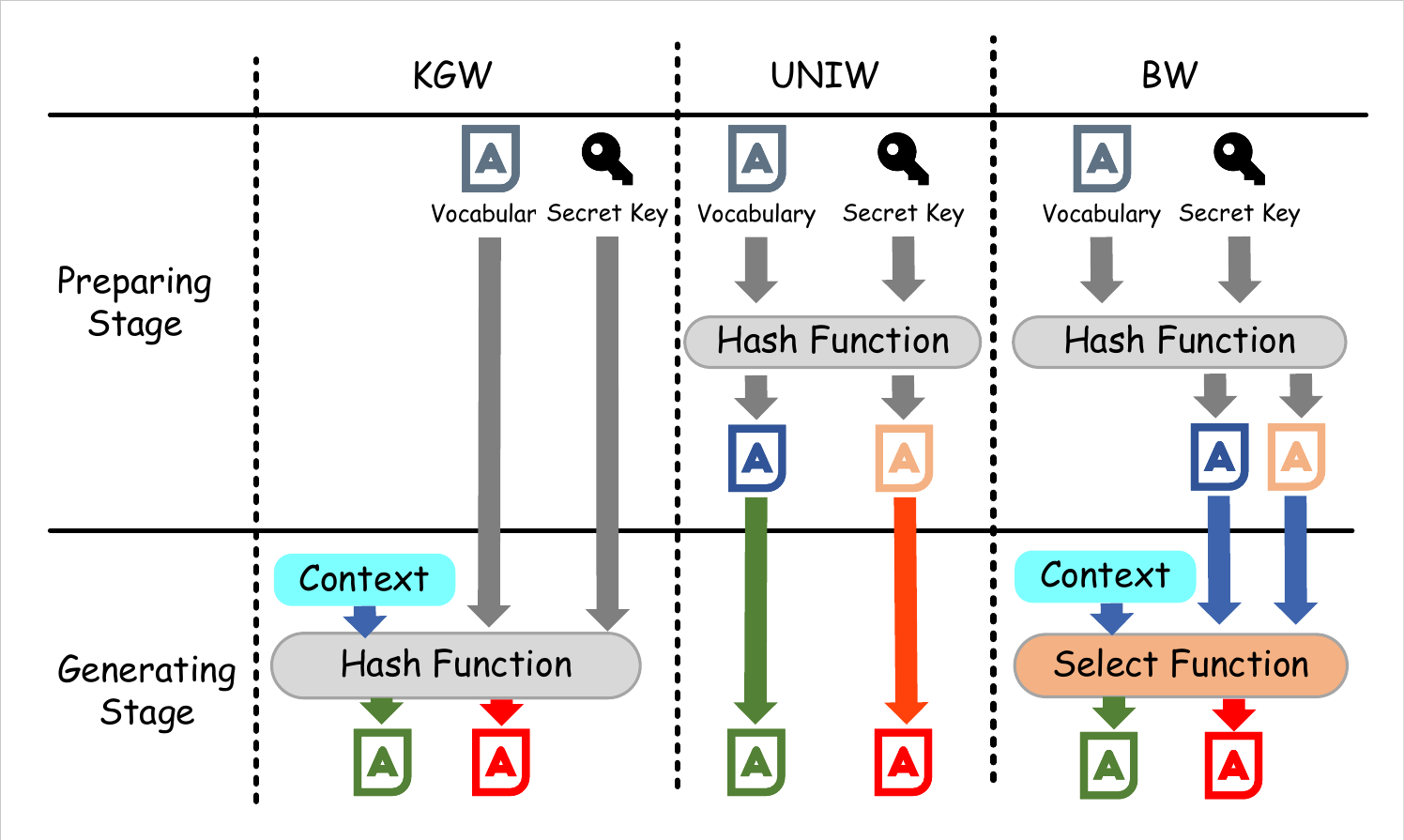} 
    \vspace{-4mm}
  \caption{Difference of partition functions in KGW, UNIW, and BW. Preparing Stage is before entering the prompt to LLM. Generating Stage is at when LLM generates a response.}
  \label{fig:BW}
\end{figure}
We show the difference in the partition function between KGW, UNIW, and BW in Fig. \ref{fig:BW}. 

KGW partition vocabulary at generating stage, the introduction of context makes $\mathcal{G}$ and $\mathcal{R}$ change at each generation step. 
UNIW partition vocabulary at the preparing stage, $\mathcal{G}$ and $\mathcal{R}$ are constant during generation. 
In the UNIW study, Zhao et al. \cite{WMUNIW23} prove that fixed $\mathcal{G}$ and $\mathcal{R}$ will bring higher robustness. 
However, fixed $\mathcal{G}$ also introduces some tokens with an exceptionally high frequency, which in turn makes spoofing attacks more feasible to execute. 

Addressing the limitation of UNIW, BW designs a Select Function to dynamically confirm $\mathcal{G}$ and $\mathcal{R}$. 
At preparing stage, BW, like UNIW, utilizes fixed list partition to obtain two subsets $\mathcal{A}$ and $\mathcal{B}$ by dividing the vocabulary to. 
At the generating stage, BW inputs context into Select Function, which then determines $\mathcal{G}$ by choosing between $\mathcal{A}$ and $\mathcal{B}$.  
$\mathcal{G}$ is dynamic for BW, just as for KGW. 

When generating the token $y_i$, the context used for the watermark is $\{y_{i-w}, y_{i-w+1}, ..., y_{i-1}\}$, $w$ denotes the length of the context window, which can be seen as watermark complexity. 
The Select Function $F_S$ maps the context to a signal and will determine which of $\mathcal{A}$ or $\mathcal{B}$ is determined as $\mathcal{G}$. 
The process of selecting $\mathcal{G}$ can be formalized as follows:  
\begin{equation}
    \label{eq:SelectG} 
    \mathcal{G}=
    \begin{cases}
        \mathcal{A}, & F_S(\{y_{i-w}, y_{i-w+1}, ..., y_{i-1}\})=1 \\
        \mathcal{B}, & F_S(\{y_{i-w}, y_{i-w+1}, ..., y_{i-1}\})=0
    \end{cases}
\end{equation}

To avoid the unusually high frequency in tokens, the probability that $\mathcal{A}$ and $\mathcal{B}$ are selected as $\mathcal{G}$ should be approximately equal. 
BW utilizes the first token $y_{i-w}$ of context to determine $\mathcal{G}$ like KGW. 
Therefore, we introduce token frequency statistics to make the mapping of $F_S$ as balanced as possible. 
BW counts the frequency of each token $t$ in $\mathcal{V}$ across a large set of non-watermarked texts generated by the LLM.
Based on the token frequency, we form an ordered list $\{t_1, t_2, ..., t_{|V|}\}$. 
Following this ordered list, when $t_j$ is $y_{i-w}$, the first token of context , $F_S$ can be formalized as:
\begin{equation}
    \label{eq:SelectFunction}
    F_S(t_j)=\begin{cases}
        1,  & j\%2=0 \\
        0,  & j\%2\neq 0
    \end{cases}
\end{equation}

We present the overall process for BW in Appendix \ref{app:BW}.

\section{Experiments} 
\begin{table*}[]
\centering
\caption{The result of different watermarks under CEFW. $S_D$, $S_T$, $S_U$, $S_I$, and $S_R$ are respectively the scores of detectability, text quality, usability, imperceptibility, and robustness after normalization under CEFW. $S_{CEFW}$ represents the comprehensive score of watermark under CEFW. Darker colors indicate better relative rankings.}
\label{tab:comprehensive}
\begin{tabular}{
>{\columncolor[HTML]{FFFFFF}}c 
>{\columncolor[HTML]{FFFFFF}}c 
>{\columncolor[HTML]{FFFFFF}}c 
>{\columncolor[HTML]{FFFFFF}}c 
>{\columncolor[HTML]{FFFFFF}}c 
>{\columncolor[HTML]{FFFFFF}}c 
>{\columncolor[HTML]{FFFFFF}}c 
>{\columncolor[HTML]{FFFFFF}}c 
>{\columncolor[HTML]{FFFFFF}}c 
>{\columncolor[HTML]{FFFFFF}}c 
>{\columncolor[HTML]{FFFFFF}}c 
>{\columncolor[HTML]{FFFFFF}}c }
                                                     &                                                    &            & UNIW  & KGW1  & KGW2  & KGW3  & KGW4  & BW1            & BW2   & BW3   & BW4                           \\ \hline
\cellcolor[HTML]{FFFFFF}                             & \cellcolor[HTML]{FFFFFF}                           & $S_D$      & \cellcolor[HTML]{FFEFEF}0.998 & \cellcolor[HTML]{FFA6A6}1.000 & \cellcolor[HTML]{FFA6A6}1.000 & \cellcolor[HTML]{FFCDCD}0.999 & \cellcolor[HTML]{FFEFEF}0.998 & \cellcolor[HTML]{FFCDCD}0.999          & \cellcolor[HTML]{FFEFEF}0.998 & \cellcolor[HTML]{FFA6A6}1.000 & \cellcolor[HTML]{FFCDCD}0.999                         \\

\cellcolor[HTML]{FFFFFF}                             & \cellcolor[HTML]{FFFFFF}                           & $S_T$      & \cellcolor[HTML]{FF9999}0.571 & \cellcolor[HTML]{FFCDCD}0.454 & \cellcolor[HTML]{FFE3E3}0.403 & \cellcolor[HTML]{FFEFEF}0.322 & \cellcolor[HTML]{FFFFFF}0.287 & \cellcolor[HTML]{FFA6A6}0.527          & \cellcolor[HTML]{FFB3B3}0.510 & \cellcolor[HTML]{FFC0C0}0.466 & \cellcolor[HTML]{FFD8D8}0.443                         \\

\cellcolor[HTML]{FFFFFF}                             & \cellcolor[HTML]{FFFFFF}                           & $S_U$      & \cellcolor[HTML]{FF9999}0.953 & \cellcolor[HTML]{FFE3E3}0.937 & \cellcolor[HTML]{FFD8D8}0.944 & \cellcolor[HTML]{FFFFFF}0.925 & \cellcolor[HTML]{FFFFFF}0.925 & \cellcolor[HTML]{FFC0C0}0.948          & \cellcolor[HTML]{FFD8D8}0.944 & \cellcolor[HTML]{FF9999}0.953 & \cellcolor[HTML]{FFC0C0}0.948                         \\

\cellcolor[HTML]{FFFFFF}                             & \cellcolor[HTML]{FFFFFF}                           & $S_I$      & \cellcolor[HTML]{FFFFFF}0.000 & \cellcolor[HTML]{FFFFFF}0.000 & \cellcolor[HTML]{FFD8D8}0.007 & \cellcolor[HTML]{FFC0C0}0.297 & \cellcolor[HTML]{FF9999}0.678 & \cellcolor[HTML]{FFE3E3}0.003          & \cellcolor[HTML]{FFCDCD}0.018 & \cellcolor[HTML]{FFB3B3}0.301 & \cellcolor[HTML]{FFA6A6}0.664                         \\

\cellcolor[HTML]{FFFFFF}                             & \cellcolor[HTML]{FFFFFF}                           & $S_R$      & \cellcolor[HTML]{FF9999}0.986 & \cellcolor[HTML]{FFB3B3}0.647 & \cellcolor[HTML]{FFCDCD}0.534 & \cellcolor[HTML]{FFE3E3}0.459 & \cellcolor[HTML]{FFFFFF}0.355 & \cellcolor[HTML]{FFA6A6}0.673          & \cellcolor[HTML]{FFC0C0}0.571 & \cellcolor[HTML]{FFD8D8}0.466 & \cellcolor[HTML]{FFEFEF}0.443             \\             
\noalign{\vskip -2pt} \cmidrule[0.4pt]{3-12} \noalign{\vskip -3pt}

\cellcolor[HTML]{FFFFFF}                             & \multirow{-6}{*}{\cellcolor[HTML]{FFFFFF}C4}       & $S_{CEFW}$ & \cellcolor[HTML]{FFA6A6}0.667 & \cellcolor[HTML]{FFE3E3}0.560 & \cellcolor[HTML]{FFFFFF}0.526 & \cellcolor[HTML]{FFD8D8}0.563 & \cellcolor[HTML]{FFB3B3}0.626 & \cellcolor[HTML]{FFCDCD}0.581          & \cellcolor[HTML]{FFEFEF}0.556 & \cellcolor[HTML]{FFC0C0}0.595 & \cellcolor[HTML]{FF9999}\textbf{0.675}                \\ \noalign{\vskip -2pt} \cmidrule[0.4pt]{2-12} \noalign{\vskip -3pt}

\cellcolor[HTML]{FFFFFF}                             & \cellcolor[HTML]{FFFFFF}                           & $S_D$      & \cellcolor[HTML]{FFCDCD}1.000 & \cellcolor[HTML]{FFCDCD}1.000 & \cellcolor[HTML]{FFCDCD}1.000 & \cellcolor[HTML]{FFCDCD}1.000 & \cellcolor[HTML]{FFCDCD}1.000 & \cellcolor[HTML]{FFCDCD}1.000          & \cellcolor[HTML]{FFCDCD}1.000 & \cellcolor[HTML]{FFCDCD}1.000 & \cellcolor[HTML]{FFCDCD}1.000                         \\

\cellcolor[HTML]{FFFFFF}                             & \cellcolor[HTML]{FFFFFF}                           & $S_T$      & \cellcolor[HTML]{FFFFFF}0.345 & \cellcolor[HTML]{FFEFEF}0.385 & \cellcolor[HTML]{FFE3E3}0.388 & \cellcolor[HTML]{FFCDCD}0.457 & \cellcolor[HTML]{FFD8D8}0.453 & \cellcolor[HTML]{FF9999}0.671          & \cellcolor[HTML]{FFA6A6}0.528 & \cellcolor[HTML]{FFB3B3}0.479 & \cellcolor[HTML]{FFC0C0}0.477                         \\

\cellcolor[HTML]{FFFFFF}                             & \cellcolor[HTML]{FFFFFF}                           & $S_U$      & \cellcolor[HTML]{FFB3B3}0.992 & \cellcolor[HTML]{FFFFFF}0.980 & \cellcolor[HTML]{FFFFFF}0.980 & \cellcolor[HTML]{FFD8D8}0.982 & \cellcolor[HTML]{FFFFFF}0.980 & \cellcolor[HTML]{FFA6A6}0.995          & \cellcolor[HTML]{FFCDCD}0.989 & \cellcolor[HTML]{FFC0C0}0.991 & \cellcolor[HTML]{FFA6A6}0.995                         \\

\cellcolor[HTML]{FFFFFF}                             & \cellcolor[HTML]{FFFFFF}                           & $S_I$      & \cellcolor[HTML]{FFFFFF}0.000 & \cellcolor[HTML]{FFFFFF}0.000 & \cellcolor[HTML]{FFE3E3}0.002 & \cellcolor[HTML]{FFB3B3}0.104 & \cellcolor[HTML]{FFA6A6}0.388 & \cellcolor[HTML]{FFE3E3}0.002          & \cellcolor[HTML]{FFCDCD}0.006 & \cellcolor[HTML]{FFC0C0}0.098 & \cellcolor[HTML]{FF9999}0.438                         \\

\cellcolor[HTML]{FFFFFF}                             & \cellcolor[HTML]{FFFFFF}                           & $S_R$      & \cellcolor[HTML]{FF9999}0.930 & \cellcolor[HTML]{FFB3B3}0.790 & \cellcolor[HTML]{FFC0C0}0.734 & \cellcolor[HTML]{FFEFEF}0.516 & \cellcolor[HTML]{FFFFFF}0.408 & \cellcolor[HTML]{FFA6A6}0.874          & \cellcolor[HTML]{FFCDCD}0.730 & \cellcolor[HTML]{FFD8D8}0.626 & \cellcolor[HTML]{FFE3E3}0.552                         \\ \noalign{\vskip -2pt} \cmidrule[0.4pt]{3-12} \noalign{\vskip -3pt}

\multirow{-12}{*}{\cellcolor[HTML]{FFFFFF}OPT-2.7b}  & \multirow{-6}{*}{\cellcolor[HTML]{FFFFFF}Quora-QA} & $S_{CEFW}$ & \cellcolor[HTML]{FFB3B3}0.622 & \cellcolor[HTML]{FFE3E3}0.592 & \cellcolor[HTML]{FFEFEF}0.579 & \cellcolor[HTML]{FFFFFF}0.561 & \cellcolor[HTML]{FFC0C0}0.604 & \cellcolor[HTML]{FF9999}\textbf{0.663} & \cellcolor[HTML]{FFCDCD}0.603 & \cellcolor[HTML]{FFD8D8}0.593 & \cellcolor[HTML]{FFA6A6}0.659 \\ \hline

\cellcolor[HTML]{FFFFFF}                             & \cellcolor[HTML]{FFFFFF}                           & $S_D$      & \cellcolor[HTML]{FFE3E3}0.997 & \cellcolor[HTML]{FFE3E3}0.997 & \cellcolor[HTML]{FFE3E3}0.997 & \cellcolor[HTML]{FFFFFF}0.995 & \cellcolor[HTML]{FFA6A6}0.999 & \cellcolor[HTML]{FFCDCD}0.998          & \cellcolor[HTML]{FFA6A6}0.999 & \cellcolor[HTML]{FFA6A6}0.999 & \cellcolor[HTML]{FFCDCD}0.998                         \\

\cellcolor[HTML]{FFFFFF}                             & \cellcolor[HTML]{FFFFFF}                           & $S_T$      & \cellcolor[HTML]{FF9999}0.694 & \cellcolor[HTML]{FFA6A6}0.602 & \cellcolor[HTML]{FFCDCD}0.564 & \cellcolor[HTML]{FFFFFF}0.512 & \cellcolor[HTML]{FFE3E3}0.525 & \cellcolor[HTML]{FFB3B3}0.582          & \cellcolor[HTML]{FFC0C0}0.576 & \cellcolor[HTML]{FFEFEF}0.524 & \cellcolor[HTML]{FFD8D8}0.537                         \\

\cellcolor[HTML]{FFFFFF}                             & \cellcolor[HTML]{FFFFFF}                           & $S_U$      & \cellcolor[HTML]{FF9999}0.998 & \cellcolor[HTML]{FFEFEF}0.980 & \cellcolor[HTML]{FFFFFF}0.979 & \cellcolor[HTML]{FFD8D8}0.982 & \cellcolor[HTML]{FFE3E3}0.981 & \cellcolor[HTML]{FFC0C0}0.992          & \cellcolor[HTML]{FFC0C0}0.992 & \cellcolor[HTML]{FFA6A6}0.993 & \cellcolor[HTML]{FFC0C0}0.992                         \\

\cellcolor[HTML]{FFFFFF}                             & \cellcolor[HTML]{FFFFFF}                           & $S_I$      & \cellcolor[HTML]{FFEFEF}0.002 & \cellcolor[HTML]{FFFFFF}0.000 & \cellcolor[HTML]{FFD8D8}0.022 & \cellcolor[HTML]{FFC0C0}0.388 & \cellcolor[HTML]{FFA6A6}0.786 & \cellcolor[HTML]{FFEFEF}0.002          & \cellcolor[HTML]{FFCDCD}0.032 & \cellcolor[HTML]{FFB3B3}0.436 & \cellcolor[HTML]{FF9999}0.768                         \\

\cellcolor[HTML]{FFFFFF}                             & \cellcolor[HTML]{FFFFFF}                           & $S_R$      & \cellcolor[HTML]{FFA6A6}0.772 & \cellcolor[HTML]{FFCDCD}0.583 & \cellcolor[HTML]{FFD8D8}0.554 & \cellcolor[HTML]{FFE3E3}0.515 & \cellcolor[HTML]{FFFFFF}0.451 & \cellcolor[HTML]{FF9999}0.792          & \cellcolor[HTML]{FFB3B3}0.652 & \cellcolor[HTML]{FFC0C0}0.588 & \cellcolor[HTML]{FFEFEF}0.502                         \\ \noalign{\vskip -2pt} \cmidrule[0.4pt]{3-12} \noalign{\vskip -3pt}

\cellcolor[HTML]{FFFFFF}                             & \multirow{-6}{*}{\cellcolor[HTML]{FFFFFF}C4}       & $S_{CEFW}$ & \cellcolor[HTML]{FFCDCD}0.641 & \cellcolor[HTML]{FFEFEF}0.576 & \cellcolor[HTML]{FFFFFF}0.567 & \cellcolor[HTML]{FFCDCD}0.641 & \cellcolor[HTML]{FFA6A6}0.727 & \cellcolor[HTML]{FFD8D8}0.627          & \cellcolor[HTML]{FFE3E3}0.599 & \cellcolor[HTML]{FFB3B3}0.675 & \cellcolor[HTML]{FF9999}\textbf{0.739}                \\ \noalign{\vskip -2pt} \cmidrule[0.4pt]{2-12} \noalign{\vskip -3pt}

\cellcolor[HTML]{FFFFFF}                             & \cellcolor[HTML]{FFFFFF}                           & $S_D$      & \cellcolor[HTML]{FFFFFF}0.998 & \cellcolor[HTML]{FFC0C0}1.000 & \cellcolor[HTML]{FFEFEF}0.999 & \cellcolor[HTML]{FFC0C0}1.000 & \cellcolor[HTML]{FFC0C0}1.000 & \cellcolor[HTML]{FFEFEF}0.999          & \cellcolor[HTML]{FFC0C0}1.000 & \cellcolor[HTML]{FFC0C0}1.000 & \cellcolor[HTML]{FFC0C0}1.000                         \\

\cellcolor[HTML]{FFFFFF}                             & \cellcolor[HTML]{FFFFFF}                           & $S_T$      & \cellcolor[HTML]{FF9999}0.781 & \cellcolor[HTML]{FFA6A6}0.651 & \cellcolor[HTML]{FFCDCD}0.589 & \cellcolor[HTML]{FFEFEF}0.534 & \cellcolor[HTML]{FFFFFF}0.453 & \cellcolor[HTML]{FFB3B3}0.660          & \cellcolor[HTML]{FFC0C0}0.625 & \cellcolor[HTML]{FFD8D8}0.578 & \cellcolor[HTML]{FFE3E3}0.545                         \\

\cellcolor[HTML]{FFFFFF}                             & \cellcolor[HTML]{FFFFFF}                           & $S_U$      & \cellcolor[HTML]{FF9999}0.996 & \cellcolor[HTML]{FFCDCD}0.982 & \cellcolor[HTML]{FFB3B3}0.983 & \cellcolor[HTML]{FFB3B3}0.983 & \cellcolor[HTML]{FFB3B3}0.983 & \cellcolor[HTML]{FFE3E3}0.981          & \cellcolor[HTML]{FFE3E3}0.981 & \cellcolor[HTML]{FFEFEF}0.980 & \cellcolor[HTML]{FFFFFF}0.952                         \\

\cellcolor[HTML]{FFFFFF}                             & \cellcolor[HTML]{FFFFFF}                           & $S_I$      & \cellcolor[HTML]{FFFFFF}0.000 & \cellcolor[HTML]{FFFFFF}0.000 & \cellcolor[HTML]{FFCDCD}0.014 & \cellcolor[HTML]{FFC0C0}0.154 & \cellcolor[HTML]{FFA6A6}0.556 & \cellcolor[HTML]{FFE3E3}0.002          & \cellcolor[HTML]{FFE3E3}0.002 & \cellcolor[HTML]{FFB3B3}0.190 & \cellcolor[HTML]{FF9999}0.586                         \\

\cellcolor[HTML]{FFFFFF}                             & \cellcolor[HTML]{FFFFFF}                           & $S_R$      & \cellcolor[HTML]{FFA6A6}0.784 & \cellcolor[HTML]{FFC0C0}0.664 & \cellcolor[HTML]{FFD8D8}0.636 & \cellcolor[HTML]{FFEFEF}0.612 & \cellcolor[HTML]{FFFFFF}0.460 & \cellcolor[HTML]{FF9999}0.812          & \cellcolor[HTML]{FFB3B3}0.732 & \cellcolor[HTML]{FFCDCD}0.648 & \cellcolor[HTML]{FFE3E3}0.620                         \\ \noalign{\vskip -2pt} \cmidrule[0.4pt]{3-12} \noalign{\vskip -3pt}

\multirow{-12}{*}{\cellcolor[HTML]{FFFFFF}Llama3-8b} & \multirow{-6}{*}{\cellcolor[HTML]{FFFFFF}Quora-QA} & $S_{CEFW}$ & \cellcolor[HTML]{FFB3B3}0.659 & \cellcolor[HTML]{FFEFEF}0.605 & \cellcolor[HTML]{FFFFFF}0.591 & \cellcolor[HTML]{FFE3E3}0.611 & \cellcolor[HTML]{FFA6A6}0.660 & \cellcolor[HTML]{FFC0C0}0.643          & \cellcolor[HTML]{FFD8D8}0.618 & \cellcolor[HTML]{FFCDCD}0.636 & \cellcolor[HTML]{FF9999}\textbf{0.718}                \\ \hline
\end{tabular}
\end{table*}

\subsection{Experiment Setup}
\label{exp:environment}

\paragraph{Dataset}
We employ C4 \cite{DSC4:journals/jmlr/RaffelSRLNMZLL20}\footnote{\url{https://huggingface.co/datasets/allenai/c4}} for our text generation task, and Quora-QA \cite{DSQuora:conf/ijcai/WangHF17}\footnote{\url{https://huggingface.co/datasets/toughdata/quora-question-answer-dataset}} to simulate the LLM service scenario. 
For input, we extract the first 30 tokens from each text in C4 and utilize the question part in Quora-QA.

\paragraph{Language Model}
We select two language models.
One is OPT-2.7b in the OPT family \cite{Modeloptzhang2022opt}, which is widely utilized for watermark evaluation. 
Another is Llama3-8b \cite{Modell3llama3modelcard}, which is a latest large language model. 
During each generation, we employ sampling as the decoding strategy and generate a maximum of 200 tokens. 

\paragraph{Evaluated Watermarks}
We compare KGW \cite{KGW} and UNIW \cite{WMUNIW23} with BW. 
For each watermark, we equally partition vocabulary and set $\delta$ as 2. 
We define KGW$w$ for KGW with watermark complexity $w$ and set watermark complexity 1 to 4 for both KGW and BW.

\subsection{Comprehensive Analysis} 

\textbf{By comparing different watermarks, BW4 is the most recommended watermark. }
Table \ref{tab:comprehensive} presents five characteristic scores and the comprehensive score for all watermarks.

The imperceptibility score of UNIW is always 0, which means STEAL can completely crack it. 
KGW with high watermark complexity does not exhibit any significant drawbacks. 
However, in the majority of instances, BW consistently demonstrates superior text quality and robustness compared to KGW when both employ the same watermark complexity. 
In the remaining cases, the scores for the corresponding characteristics are approximately equal. 
In the comparison of comprehensive scores, BW4 achieves the highest comprehensive score in three settings, and BW1 achieves in one. 
However, BW4 has no obvious shortcomings with all characteristic scores higher than 0.4, while BW1 has significant imperceptibility disadvantages. 

In our design, BW achieves a high comprehensive score for two reasons:
1) \textbf{Select Function brings higher imperceptibility than UNIW}; 
2) \textbf{Fixed list partition brings higher robustness than KGW}. 

\begin{figure*}[ht!]
    \centering
    \includegraphics[width=0.98\textwidth]{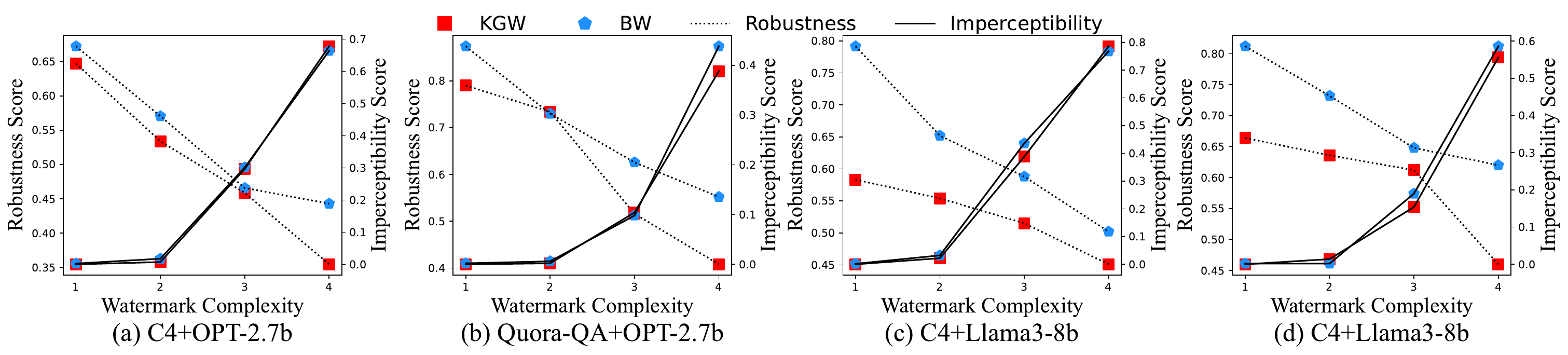}
      \vspace{-3mm}
    \caption{Watermark Complexity Analysis. }
    \label{fig:watermarkComplexity}
\end{figure*} 

Select Function enables BW to regulate its watermark complexity. 
As shown in Table \ref{tab:comprehensive}, watermark complexity significantly affects the imperceptibility. 
Watermarks with high watermark complexity (KGW3, KGW4, BW3, BW4) have imperceptibility scores higher than UNIW obviously. 
By increasing watermark complexity, BW can achieve a higher imperceptibility score compared to the UNIW. 

According to the result of Zhao et al. \cite{WMUNIW23}, fixed list partition and selection will bring the watermarks higher robustness. 
Fixed list partition makes the watermark information of BW more stubborn. 
Thus, BW exhibits robustness that is superior to or equivalent to that of KGW. 

For other characteristics, the three watermarks achieve scores greater than 0.9 in detectability and usability. 
In text quality, BW shows a more stable advantage over KGW, with scores that are either higher or equivalent. 
In contrast, the text quality of UNIW may suffer a significant decline due to an unreasonable list partition.

\subsection{Watermark Complexity Analysis}

\textbf{Randomness brings watermark higher imperceptibility but lower robustness. }

Tokens that are closer in natural language are generally considered to have stronger correlations. 
Higher watermark complexity makes the token selected in the context and the token to be generated have a lower correlation, resulting in a more random green list. 

As shown in Fig. \ref{fig:watermarkComplexity}, the additional randomness introduced by the increase in watermark complexity improves the imperceptibility of BW and KGW at the cost of reduced robustness.

\subsection{STEAL Analysis}

\textbf{Select Function brings close imperceptibility as KGW for BW.} 
To further validate it, we present the AUCROC values of texts generated by 4 STEAL attacks in Table \ref{tab:Imperceptibility}. 

We observe that a fixed configuration STEAL, when attacking BW and KGW with the same complexity, generates spoofing texts with AUCROC scores that mostly differ by no more than 0.05, exhibiting closely similar AUCROC scores. 
As a result, no matter how the watermark setting or attack setting changes, the imperceptibility brought about by Select Function for BW is always similar to that of KGW.

\begin{table*}[]
\centering
\caption{The AUCROC value of all watermarks for imperceptibility evaluation. STEAL$n$ represents the $n$-gram frequency analysis for this STEAL attack. A. represents Authorization, and N.A. represents Non-authorization. }
\label{tab:Imperceptibility}
\begin{tabular}{cccccccccccc}
                            &                           &        & UNIW  & KGW1  & KGW2  & KGW3  & KGW4  & BW1   & BW2   & BW3   & BW4   \\ \hline
\multirow{12}{*}{OPT-2.7b}  & \multirow{6}{*}{C4}       & STEAL1 & 1.000 & 1.000 & 0.874 & 0.594 & 0.551 & 0.998 & 0.891 & 0.611 & 0.546 \\
                            &                           & STEAL2 & 0.989 & 0.991 & 0.996 & 0.788 & 0.639 & 0.990 & 0.991 & 0.797 & 0.657 \\
                            &                           & STEAL3 & 0.762 & 0.838 & 0.826 & 0.852 & 0.661 & 0.814 & 0.837 & 0.849 & 0.668 \\
                            &                           & STEAL4 & 0.574 & 0.629 & 0.597 & 0.637 & 0.655 & 0.614 & 0.608 & 0.647 & 0.644 \\ \cline{3-12} 
                            &                           & N.A.   & 0.832 & 0.865 & 0.823 & 0.718 & 0.626 & 0.854 & 0.832 & 0.726 & 0.629 \\
                            &                           & A.     & 1.000 & 1.000 & 0.996 & 0.852 & 0.661 & 0.998 & 0.991 & 0.849 & 0.668 \\ \cline{2-12} 
                            & \multirow{6}{*}{Quora-QA} & STEAL1 & 1.000 & 1.000 & 0.832 & 0.661 & 0.543 & 0.999 & 0.829 & 0.669 & 0.571 \\
                            &                           & STEAL2 & 0.995 & 0.999 & 0.999 & 0.828 & 0.637 & 0.997 & 0.997 & 0.787 & 0.655 \\
                            &                           & STEAL3 & 0.903 & 0.829 & 0.930 & 0.948 & 0.730 & 0.924 & 0.936 & 0.951 & 0.683 \\
                            &                           & STEAL4 & 0.700 & 0.733 & 0.735 & 0.779 & 0.806 & 0.726 & 0.756 & 0.748 & 0.781 \\ \cline{3-12} 
                            &                           & N.A.   & 0.900 & 0.890 & 0.874 & 0.804 & 0.679 & 0.912 & 0.880 & 0.789 & 0.673 \\
                            &                           & A.     & 1.000 & 1.000 & 0.999 & 0.948 & 0.806 & 0.999 & 0.997 & 0.951 & 0.781 \\ \hline
\multirow{12}{*}{Llama3-8b} & \multirow{6}{*}{C4}       & STEAL1 & 0.999 & 1.000 & 0.768 & 0.603 & 0.563 & 0.999 & 0.676 & 0.581 & 0.533 \\
                            &                           & STEAL2 & 0.968 & 0.992 & 0.989 & 0.733 & 0.604 & 0.985 & 0.984 & 0.712 & 0.616 \\
                            &                           & STEAL3 & 0.674 & 0.786 & 0.774 & 0.806 & 0.607 & 0.726 & 0.756 & 0.782 & 0.583 \\
                            &                           & STEAL4 & 0.550 & 0.568 & 0.563 & 0.553 & 0.576 & 0.588 & 0.569 & 0.575 & 0.583 \\ \cline{3-12} 
                            &                           & N.A.   & 0.798 & 0.837 & 0.774 & 0.674 & 0.588 & 0.825 & 0.746 & 0.663 & 0.579 \\
                            &                           & A.     & 0.999 & 1.000 & 0.989 & 0.806 & 0.607 & 0.999 & 0.984 & 0.782 & 0.616 \\ \cline{2-12} 
                            & \multirow{6}{*}{Quora-QA} & STEAL1 & 1.000 & 1.000 & 0.791 & 0.584 & 0.564 & 0.999 & 0.750 & 0.626 & 0.536 \\
                            &                           & STEAL2 & 0.987 & 0.998 & 0.993 & 0.828 & 0.645 & 0.996 & 0.999 & 0.791 & 0.639 \\
                            &                           & STEAL3 & 0.821 & 0.910 & 0.914 & 0.923 & 0.695 & 0.875 & 0.901 & 0.905 & 0.663 \\
                            &                           & STEAL4 & 0.623 & 0.678 & 0.659 & 0.676 & 0.722 & 0.656 & 0.672 & 0.683 & 0.707 \\ \cline{3-12} 
                            &                           & N.A.   & 0.858 & 0.897 & 0.839 & 0.753 & 0.657 & 0.882 & 0.831 & 0.751 & 0.636 \\
                            &                           & A.     & 1.000 & 1.000 & 0.993 & 0.923 & 0.722 & 0.999 & 0.999 & 0.905 & 0.707 \\ \hline
\end{tabular}
\end{table*}

\section{Conclusion} 
In this paper, we propose a Comprehensive Evaluation Framework for Watermark in LLMs (CEFW). 
CEFW standardizes the rational evaluation process for watermarks by considering five necessary characteristics. 
We also propose a novel watermark, termed Balanced Watermark (BW), which balances the method to add watermark information. 
In the experimental part, we prove that BW has the best comprehensive performance among the three watermarks.

\section*{Acknowledgment}
This work was supported in part by the National Key Research and Development Program of China under Grant 2022YFF0902701; in part by the National Natural Science Foundation of China under Grants U21A20468, 62372058, U22A2026.

\bibliographystyle{IEEEbib}
\bibliography{conference_101719}

\vspace{12pt}

\appendix
\section{Appendix}
\subsection{Algorithm}
\subsubsection{Balanced Watermark}
\label{app:BW}
We summarize Section \ref{sec:BW} and give the overall flow algorithm of BW in Algorithm \ref{alg:BW}.

\begin{algorithm}[h]
    \begin{algorithmic}[1]
        \REQUIRE Prompt $Y=\{y_1, y_2, ...y_{|Y|}\}$, Large Language Model $LLM$, secret key $\mathcal{K}$, logits bias $\delta > 0$, watermark complexity $w$.
        \ENSURE watermarked text
        \STATE Count token frequencies from large amounts of text generated by $LLM$;
        \STATE Sort tokens on $\mathcal{V}$ by token frequencies;
        \STATE Construct Select Function $F_S$ according to the order tokens list;
        \STATE Apply $\mathcal{K}$ as a random seed, randomly and uniformly partition the vocabulary $\mathcal{V}$ into lists $\mathcal{A}$ and $\mathcal{B}$;
        \FOR{$i \leftarrow |Y|+1$ \TO ...}
            \STATE Based on prompt $y_{<i}$, LLM get a logits distribution $l^{(i)}$ on the vocabulary $\mathcal{V}$;
            \IF{$F_S(y_{i-w})=1$}
                \STATE{$\mathcal{G}=\mathcal{A}$, $\mathcal{R}=\mathcal{B}$}
            \ELSIF{$F_S(y_{i-w})=0$}
                \STATE{$\mathcal{G}=\mathcal{B}$, $\mathcal{R}=\mathcal{A}$}
            \ENDIF
            \STATE{Add a fixed bias value $\delta$ to all green tokens logits, then obtain a new probability distribution $p_{w}^{(i)}$ over the vocabulary $
            \mathcal{V}$ through softmax;}
            \STATE{Sample the next token $y_i$ from $p_{w}^{(i)}$.} 
        \ENDFOR
    \end{algorithmic}
    \caption{Balanced Watermark}
    \label{alg:BW}
\end{algorithm}
\subsubsection{KGW}
In this paper, we evaluate the most commonly used Soft Watermark of KGW. 
To be consistent with BW, we set the green list ratio to 0.5. 
The algorithm is shown in Algorithm \ref{alg:KGW}.
\begin{algorithm}[h]
    \begin{algorithmic}[1]
        \REQUIRE{Prompt $Y=\{y_1, y_2, ...y_{|Y|}\}$, Large Language Model $LLM$, secret key $\mathcal{K}$, logits bias $\delta > 0$, watermark complexity $w$.}
        \ENSURE{watermarked text.}
        \FOR{$i \leftarrow |Y|+1$ \TO ...}
            \STATE{Based on prompt $y_{<i}$, LLM get a logits distribution $l^{(i)}$ on the vocabulary $\mathcal{V}$;}
            \STATE{Compute a hash of token $y_{i-w}$, and use it to seed a random number generator;}
            \STATE{Using this random number generator to randomly and uniformly partition the vocabulary;}
            \STATE{Add a fixed bias value $\delta$ to all green tokens logits, then obtain a new probability distribution $p_{w}^{(i)}$ over the vocabulary $
            \mathcal{V}$ through softmax;}
            \STATE{Sample the next token $y_i$ from $p_{w}^{(i)}$.} 
        \ENDFOR
    \end{algorithmic}
    \caption{KGW}
    \label{alg:KGW}
\end{algorithm}

\subsubsection{UNIW}
Like KGW, UNIW also sets the green list ratio at 0.5. 
We show its process in Algorithm \ref{alg:UNIW}.

\begin{algorithm}[h]
    \begin{algorithmic}[1]
        \REQUIRE{Prompt $Y=\{y_1, y_2, ...y_{|Y|}\}$, Large Language Model $LLM$, secret key $\mathcal{K}$, logits bias $\delta > 0$, watermark complexity $w$.}
        \ENSURE{watermarked text }
        \STATE{Use secret key $\mathcal{K}$ to seed a random generator;}
        \STATE{Using this random number generator to randomly and uniformly partition the vocabulary;}
        \FOR{$i \leftarrow |Y|+1$ \TO ...}
            \STATE{Based on prompt $y_{<i}$, LLM get a logits distribution $l^{(i)}$ on the vocabulary $\mathcal{V}$;}
            \STATE{Add a fixed bias value $\delta$ to all green tokens logits, then obtain a new probability distribution $p_{w}^{(i)}$ over the vocabulary $
            \mathcal{V}$ through softmax;}
            \STATE{Sample the next token $y_i$ from $p_{w}^{(i)}$.} 
        \ENDFOR
    \end{algorithmic}
    \caption{UNIW}
    \label{alg:UNIW}
\end{algorithm}

\subsubsection{DIPPER}
DIPPER \cite{dipperLFQA:conf/nips/KrishnaSKWI23} is a language model for the scrubbing attack. 
In robustness evaluation, it rewrites the watermark text to avoid watermark detection while maintaining the meaning of the text. 
Typically, the watermark has high robustness if the watermark texts have high detectability after DIPPER. 

\subsubsection{STEAL}
We provide a simple introduction for the spoofing attack, STEAL \cite{AT-STEALDBLP:conf/icml/0001SV24}. 
We only introduce one configuration from multiple attack modes of STEAL, which considers only the n-gram scenario, it conducts a spoofing attack by statistically analyzing the n-gram frequency differences between non-watermarked text and watermarked texts. 

To record n-gram frequency statistic, we tokenize both non-watermarked and watermarked texts. 
After tokenizing the texts, we record the token sequences in a dictionary. 
For an n-gram spoofing attack, the dictionary counts $({T_{i-n},T_{i-n+1},...,T_{i-1}}, T_i)$ as the key, with the corresponding count as the value. 
$T_i$ represents the current token. 
To obtain the final target dictionary, we convert the frequency count of the dictionary into actual frequencies. 
Subsequently, the attacker will construct two dictionaries for both non-watermarked and watermarked texts. 

Due to the increased complexity of the spoofing attack, we can observe that the number of keys required for statistical analysis increases exponentially. 
In the worst-case scenario, the number of keys that n-gram spoofing attack needs to count is $|V|^{n}$, where $|V|$ is the size of the vocabulary.

We then leverage the dictionary to influence the generation. 
In this step, we need to determine which token needs to increase the probability when the LLM generates tokens.
The spoofing attack influences this step in a manner similar to KGW, by adding a certain bias to specific tokens to make them more likely to be generated. 
Unlike KGW, the spoofing attack determines how to add bias by comparing the two dictionaries.

Given the context $ctx$, A computes a score for the token $T$: 
\begin{equation}
    \label{eq:spoof}
    s(T,ctx)=\begin{cases}
        \frac{1}{2}min(\frac{\hat{p}_w(T|ctx)}{\hat{p}_b(T|ctx)}, 2), & \text{if } \frac{\hat{p}_w(T|ctx)}{\hat{p}_b(T|ctx)} \geq 1 \text{,} \\
        0,          & \text{otherwise.}
    \end{cases}
\end{equation}
$\hat{p}_b(T|ctx)$ represents the frequency of the non-watermarked text, while $\hat{p}_w(T|ctx)$ represents the frequency of the watermarked text.
This formula obtains the score for outlier high-frequency tokens, enabling the spoofing attack to determine which tokens are worth adding bias to and how much bias should be added.

In the final step, the spoofing attack multiplies the attack intensity by the score to obtain the final bias vector that will be added to the logits distribution.

\subsection{Experimental Result}

\paragraph{Detectability}

\begin{table}[]
    \centering
    \caption{The AUCROC value of all watermarks for detectability evaluation.}
    \label{tab:Detectability}
    \begin{tabular}{ccccc}
         & \multicolumn{2}{c}{OPT-2.7b} & \multicolumn{2}{c}{Llama3-8b} \\ \hline
         & C4          & Quora-QA       & C4           & Quora-QA       \\ \hline
    UNIW & 0.998       & 1.000          & 0.997        & 0.998          \\ \hline
    KGW1 & 1.000       & 1.000          & 0.997        & 1.000          \\
    KGW2 & 1.000       & 1.000          & 0.997        & 0.999          \\
    KGW3 & 0.999       & 1.000          & 0.995        & 1.000          \\
    KGW4 & 0.998       & 1.000          & 0.999        & 1.000          \\ \hline
    BW1  & 0.999       & 1.000          & 0.998        & 0.999          \\
    BW2  & 0.998       & 1.000          & 0.999        & 1.000          \\
    BW3  & 1.000       & 1.000          & 0.999        & 1.000          \\
    BW4  & 0.999       & 1.000          & 0.998        & 1.000          \\ \hline
    \end{tabular}
\end{table}
For four generation environments, we generate 5000 watermarked texts and 5000 non-watermarked texts and each text has at most 200 max new tokens. 
For details of detectability evaluation, we show the AUCROC values in Table \ref{tab:Detectability}.

\paragraph{Text Quality}
\begin{table}[]
\centering
\caption{The Perplexity value of all watermarks for text quality evaluation. Original means when generating without adding any watermark.}
\label{tab:TextQuality}
\begin{tabular}{ccccc}
         & \multicolumn{2}{c}{OPT-2.7b} & \multicolumn{2}{c}{Llama3-8b} \\ \hline
         & C4          & Quora-QA       & C4           & Quora-QA       \\ \hline
Original & 4.388       & 5.054          & 3.065        & 2.668          \\ \hline
UNIW     & 6.273       & 8.362          & 4.004        & 3.251          \\ \hline
KGW1     & 6.784       & 8.164          & 4.284        & 3.598          \\
KGW2     & 7.007       & 8.146          & 4.401        & 3.764          \\
KGW3     & 7.363       & 7.799          & 4.561        & 3.911          \\
KGW4     & 7.519       & 7.82           & 4.521        & 3.966          \\ \hline
BW1      & 6.464       & 6.715          & 4.345        & 3.576          \\
BW2      & 6.539       & 7.441          & 4.365        & 3.669          \\
BW3      & 6.732       & 7.686          & 4.524        & 3.794          \\
BW4      & 6.832       & 7.698          & 4.483        & 3.882          \\ \hline
\end{tabular}
\end{table}
Perplexity is selected as the text quality evaluation metric in CEFW, and Llama-2-13b\footnote{\url{https://huggingface.co/meta-llama/Llama-2-13b-hf}} is the auxiliary language model. 
We utilize the same batch of data as for the detectability evaluation in the text quality evaluation and show the perplexity value of all watermarks in Table \ref{tab:TextQuality}. 

\paragraph{Usability} 
\begin{table*}[]
\centering
\caption{The metric values of all watermarks for usability evaluation. $t_g$ is the time to generate 5000 texts; $t_d$ is the time to detect 5000 texts; $Mem$ is the metric of Memory Cost. Original means when generating without adding any watermark. }
\label{tab:Usability}
\begin{tabular}{ccccccccccccc}
                           &                           &       & Original & UNIW    & KGW1    & KGW2    & KGW3    & KGW4    & BW1     & BW2     & BW3     & BW4     \\ \hline
\multirow{6}{*}{OPT-2.7b}  & \multirow{3}{*}{C4}       & $t_g$ & 18779    & 21379   & 22186   & 21488   & 20952   & 22038   & 21356   & 21515   & 21026   & 21256   \\
                           &                           & $t_d$ & -        & 20.79   & 216.05  & 231.77  & 263.40  & 252.01  & 90.04   & 105.07  & 112.97  & 115.15  \\
                           &                           & $Mem$ & 5057.7   & 5057.9  & 5057.9  & 5057.9  & 5057.9  & 5057.9  & 5058.3  & 5058.3  & 5058.3  & 5058.3  \\ \cline{2-13} 
                           & \multirow{3}{*}{Quora-QA} & $t_g$ & 20464    & 20862   & 20825   & 20861   & 20768   & 20878   & 20420   & 20811   & 20699   & 20298   \\
                           &                           & $t_d$ & -        & 19.48   & 204.41  & 200.07  & 196.57  & 198.58  & 82.08   & 81.62   & 81.60   & 80.31   \\
                           &                           & $Mem$ & 5057.7   & 5057.9  & 5057.9  & 5057.9  & 5057.9  & 5057.9  & 5058.3  & 5058.3  & 5058.3  & 5058.3  \\ \hline
\multirow{6}{*}{Llama3-8b} & \multirow{3}{*}{C4}       & $t_g$ & 46342    & 45975   & 45611   & 45655   & 45724   & 46157   & 45648   & 46232   & 45803   & 45611   \\
                           &                           & $t_d$ & -        & 33.80   & 299.45  & 309.03  & 275.47  & 281.09  & 124.27  & 123.29  & 110.08  & 117.67  \\
                           &                           & $Mem$ & 15316.5  & 15317.0 & 15317.0 & 15317.0 & 15317.0 & 15317.0 & 15318.1 & 15318.1 & 15318.1 & 15318.1 \\ \cline{2-13} 
                           & \multirow{3}{*}{Quora-QA} & $t_g$ & 45603    & 45978   & 46197   & 46063   & 45987   & 45973   & 47330   & 47304   & 47418   & 51258   \\
                           &                           & $t_d$ & -        & 18.85   & 209.41  & 206.15  & 208.32  & 208.67  & 96.69   & 98.32   & 98.41   & 94.84   \\
                           &                           & $Mem$ & 15316.5  & 15317.0 & 15317.0 & 15317.0 & 15317.0 & 15317.0 & 15318.1 & 15318.1 & 15318.1 & 15318.1 \\ \hline
\end{tabular}
\end{table*} 
Three metrics are selected for usability. 
For time aspect, two metrics are the time needed to detect 5000 texts and the time to generate 5000 texts. 
For memory aspect, we sum the size of the language model and the size of the additional variable introduced by the watermark as the metric of Memory Cost. 
For example, OPT-2.7b is 5057.7MB and logits bias vector from KGW is 0.191MB. 
Therefore, $M_{MemoryCost}(Y)$ is 5057.7 and $M_{MemoryCost}(\hat{Y})$ is 5057.891. 
The result of three metrics for usability is shown in Table \ref{tab:Usability}. 

\paragraph{Robustness}
\begin{table*}[]
\centering
\caption{The AUCROC value of all watermarks for robustness evaluation. No Attack represents that watermark do not suffer any scrubbing attack. DIPPER represents that the watermark has been attacked by DIPPER. }
\label{tab:Robust}
\begin{tabular}{cccccccccccc}
                           &                           &           & UNIW  & KGW1  & KGW2  & KGW3  & KGW4  & BW1   & BW2   & BW3   & BW4   \\ \hline
\multirow{4}{*}{OPT-2.7b}  & \multirow{2}{*}{C4}       & No Attack & 0.999 & 0.985 & 0.983 & 0.990 & 0.992 & 0.978 & 0.996 & 0.997 & 0.993 \\
                           &                           & DIPPER    & 0.992 & 0.814 & 0.758 & 0.720 & 0.674 & 0.822 & 0.783 & 0.729 & 0.719 \\ \cline{2-12} 
                           & \multirow{2}{*}{Quora-QA} & No Attack & 1.000 & 1.000 & 1.000 & 1.000 & 1.000 & 1.000 & 1.000 & 1.000 & 1.000 \\
                           &                           & DIPPER    & 0.965 & 0.895 & 0.867 & 0.758 & 0.704 & 0.937 & 0.865 & 0.813 & 0.776 \\ \cline{2-12} 
\multirow{4}{*}{Llama3-8b} & \multirow{2}{*}{C4}       & No Attack & 0.999 & 0.999 & 1.000 & 0.999 & 0.999 & 0.999 & 1.000 & 1.000 & 1.000 \\
                           &                           & DIPPER    & 0.885 & 0.791 & 0.777 & 0.757 & 0.725 & 0.895 & 0.826 & 0.794 & 0.751 \\ \cline{2-12} 
                           & \multirow{2}{*}{Quora-QA} & No Attack & 1.000 & 1.000 & 1.000 & 1.000 & 1.000 & 1.000 & 1.000 & 1.000 & 1.000 \\
                           &                           & DIPPER    & 0.892 & 0.832 & 0.818 & 0.806 & 0.730 & 0.906 & 0.866 & 0.824 & 0.810 \\ \hline
\end{tabular}
\end{table*}

We show the AUCROC values before and after the DIPPER attack in Table \ref{tab:Robust}. 
The AUCROC value without attack is different from the value in detectability, because we only select 500 texts to attack in robustness, but 5000 for detectability. 

\paragraph{Imperceptibility} 
For each watermark, we generate 5000 watermarked texts and 5000 non-watermarked texts for n-gram frequency statistics. 
After learning the watermark, each STEAL attack generates 500 spoofing texts. 
In the end, we calculate their AUCROC to evaluate imperceptibility.

We show our result of the imperceptibility evaluation in Table \ref{tab:Imperceptibility}.

\subsection{Flexible Design}
\label{app:CEFW}
\begin{figure}[t]
  \centering
  \includegraphics[width=0.95\columnwidth]{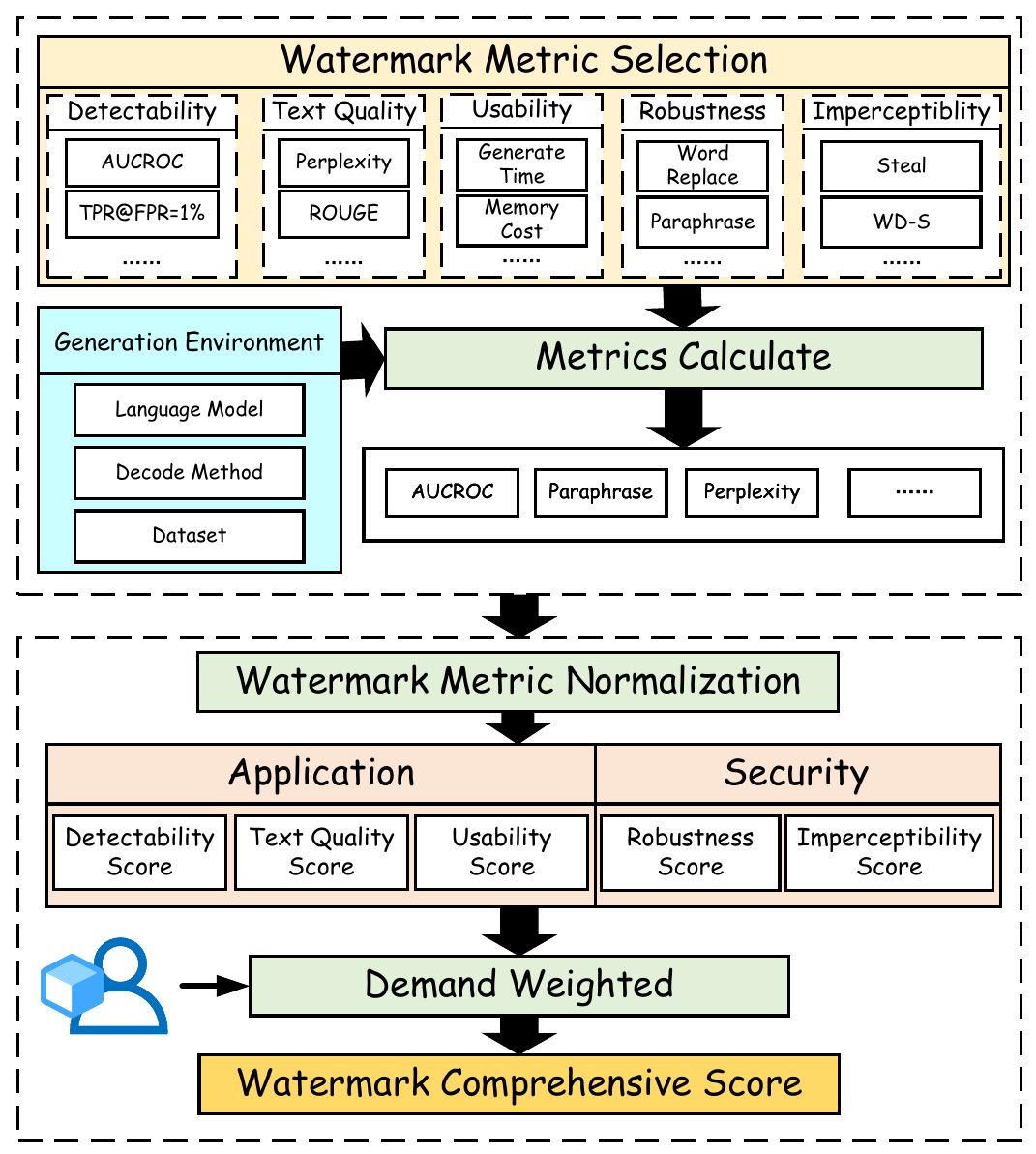} 
  \caption{An overview of CEFW. The upper area delimited by dotted lines is traditional watermark evaluation process, which is used by current watermark study. The bottom area shows the work of CEFW. CEFW combs the current watermark metrics to give five necessary characteristics score. Subsequently, CEFW introduces the Demand Weighted from LLM service providers to get a watermark comprehensive score.}
  \label{fig:CEFW}
\end{figure}
As shown in Fig. \ref{fig:CEFW}, CEFW can flexibly set the selection of watermark metrics and demand weights to better provide a suitable evaluation score for LLM service providers.
CEFW first conducts a traditional watermark evaluation process to obtain some available metric values. 
Subsequently, CEFW classifies the corresponding characteristic and normalizes all metric values to ensure that each characteristic has one final score. 
Finally, CEFW utilizes simple weighting to bring in the actual needs of the LLM service provider. 





We give some existing watermark metrics in Table \ref{tab:metrics} and associate them with the corresponding watermark characteristics. 

\begin{table}[]
\centering
\caption{Watermark Evaluation Metrics}
\label{tab:metrics}
\begin{tabular}{|c|l|}
\hline
\multirow{2}{*}{Detectability}                     & TPR/TNR/FPR/FNR; AUCROC; Accuracy; \\
                                & TPR@FPR=X\%; Bit Accuracy; Bit Error Rate.                                                                      \\ \hline
Usability                         & Generate Time; Detect Time; Memory Cost.                                                                                                                                                                                                                                            \\ \hline
\multirow{2}{*}{Text Quality}                       & Perplexity; BLEU; ROUGE; BERT Score; \\ 
& Entailment Score; Log Diversity.                                                                                                                                                                                                                \\ \hline
Robustness                        & \begin{tabular}[c]{@{}l@{}}Emoji Prompt; Word (Insert/Delete/Exchange); \\ Replace-Synonym; Replace-Context; \\ HELM Perturbation; Human Modify; \\Paraphrase; Bigram-Paraphrase \cite{WMsemstamp23}; \\ Back-translate; Re-watermark.\end{tabular} \\ \hline
\multirow{1}{*}{Imperceptibility} & Steal \cite{AT-STEALDBLP:conf/icml/0001SV24}; 181-greenList \cite{AT-spoofingDBLP:journals/corr/abs-2303-11156}; WD-S \cite{AT-LearnableDBLP:conf/iclr/GuLLH24}.                                                                                                                                                                                                                    \\ \hline
\end{tabular}
\end{table}

Although we can associate metrics with the corresponding watermark characteristics, LLM service providers cannot integrate different metrics due to their great difference in numerical feature. 
To obtain a comprehensive score, CEFW \textbf{designs some normalization functions for watermark metrics} to convert them into values in the range of 0 to 1. 
The consistent numerical features enable CEFW to integrate different metrics.
Therefore, a comprehensive score of the watermark becomes computable. 

The key to normalization is to set the upper and lower bounds. 
Give the upper bound $V_u$ and lower bound $V_l$, for a metric value $V$, the normalization result $\hat{V}$ is:
\begin{equation}
    \label{eq:normalize}
    \hat{V} = min(0, max(1, \frac{V-V_l}{V_u - V_l}))
\end{equation}
Therefore, CEFW achieves different normalization methods by setting upper and lower bounds according to the following three principles: 
\paragraph{Original Bounds}
\label{normal:original}
These metrics have their own upper and lower bounds, CEFW can directly normalize them. 
For example, AUCROC has upper bound 1 and lower bound 0.5. 
In watermark evaluation, these metrics are usually binary classification metrics for detectability. 
It is worth noting that the evaluation of imperceptibility is achieved by detecting cracked text generated by spoofing attacks, therefore the metric of imperceptibility can also utilize them. 

\paragraph{Preset Bounds}
\label{normal:preset}
Preset Bounds sets the upper and lower bounds for metrics by given values from LLM service providers or other evaluators. 
Theoretically, each metric supports Preset Bounds to achieve normalization. 
We take an example Detect Time, it is a metric that can only utilize Preset Bounds. 
CEFW regards it as the time to detect a piece of text and sets the upper bound to 0 seconds and the lower bound to 1 second. 

\paragraph{Comparison Bounds} 
\label{normal:compare}
The upper and lower bounds of metrics suitable for Comparison Bounds are usually obtained by calculating the metric value of two target texts to be compared. 

Robustness is a classic example. 
We calculate some detectability metrics to evaluate the original watermarked text and the watermarked text after scrubbing attack for robustness. 
The upper bound is set to the detectability metric value of the original watermark text, while the lower bound is the same as the lower bound of the metric. 

Sometimes, the metric to compare has no lower bound, such as Perplexity, Generate Time, and Memory Cost. 
They only have upper bounds, which are the metric values of the generated text without watermark. 
Since the generation environment affects their upper bound values, it is unreasonable to assume a fixed lower bound for them. 
CEWF sets the lower bound as twice their upper bound value, indicating that the addition of the watermark causes the worst deterioration to be twice as bad.

\end{document}